\documentclass[twocolumn,twoside,slac_two]{revtex4}
\usepackage{graphicx}
\usepackage{fancyhdr}
\def\DZero{D0~}

\newcommand{\met}{\mbox{$\raisebox{.3ex}{$\not\!$}E_T$}}
\newcommand{\fb}{$fb^{-1}$}

\begin{document}

\title{Diboson measurements at the Tevatron}

%

\author{M. Hurwitz}
\affiliation{Enrico Fermi Institute, University of Chicago, Chicago, IL 60637, USA}

\begin{abstract}
We present a summary of recent $WW$, $WZ$, and $ZZ$ measurements
carried out in $\sqrt{s} = 1.96$ TeV at the Tevatron.  The observation
of rare $ZZ$ events and the precise measurement of the $WW$ cross
section in fully leptonic decay modes are described.  Results in
semi-leptonic decay modes where one boson decays to two quarks are
also presented.  The measurements described are all in good agreement
with the Standard Model and are relevant to searches for the Higgs
boson at the Tevatron.
\end{abstract}

\maketitle

\thispagestyle{fancy}


\section{Introduction}

Measurements of diboson production cross sections at the Tevatron test
the electroweak sector of the Standard Model and have been used to
place limits on models of physics beyond the Standard
Model~\cite{WW_D0}.  Diboson measurements are also useful in the
context of searches for the Standard Model Higgs boson at the
Tevatron.  In this presentation we focus on recent diboson results
that are relevant to the Higgs searches.

The search for the Higgs boson at the Tevatron involves searching for
a very small signal in overwhelming backgrounds.  Sophisticated
analysis techniques are often used to exploit small differences
between signal and background events.  The searches also gain power
from increasing signal acceptance and dividing events into several
regions depending on their signal-to-background ratios.  Some of the
diboson searches and measurements presented below take advantage of
similar techniques while others use somewhat simpler strategies.
Comparison of the results derived with different techniques is a
useful test of the analysis strategies employed in the Higgs searches.

The measurements presented here are performed in $p\bar{p}$ collision
data with $\sqrt{s} = 1.96$~TeV collected by either the CDF II or
\DZero detector.  The detectors are described in detail
elsewhere~\cite{CDFdet}~\cite{D0det}.

\section{Fully Leptonic Decay Channels}

$WW$, $WZ$, and $ZZ$ production have all been observed at the Tevatron
at the 5$\sigma$ level in events where both bosons decay leptonically.
Table~\ref{tab:lep_sum} shows recent measurements of the cross
sections for each of the two experiments.  The measured cross sections
agree well between the two experiments and with the Standard Model
predictions.  The observation of $ZZ$ production and the measurement
of the $WW$ production cross section are discussed in more detail below.

\begin{table}
\begin{center}
\caption{\label{tab:lep_sum}Summary of diboson cross sections measured in leptonic channels at the CDF and \DZero detectors.}
\begin{tabular}{|l|c|c|c|}
\hline
\hline
 & \multicolumn{3}{c|}{Cross section [pb]} \\
Process & CDF & \DZero & NLO prediction \\
\hline
$WW$ & $12.1^{+1.8}_{-1.7}$ & $11.4 \pm 2.2$ & $11.7 \pm 0.7$ \\
$WZ$ & $4.3^{+1.3}_{-1.1}$ & $2.7^{+1.7}_{-1.3}$ & $3.7 \pm 0.3$ \\
$ZZ$ & $1.4^{+0.7}_{-0.6}$ & $1.6 \pm 0.65$ & $1.4\pm0.1$ \\
\hline
\hline
\end{tabular}
\end{center}
\end{table}

\subsection{$ZZ$ Observation}

$ZZ$ production is predicted to have a low cross section in the
Standard Model, $\sigma(p\bar{p} \rightarrow ZZ) = 1.4 \pm 0.1$~pb,
making it one of the rarest processes observed at the Tevatron so far.
Both the \DZero and CDF measurements combine a search in a four-lepton
final state ($ZZ \rightarrow llll$) with a search in a final state
with two leptons and $\met$ ($ZZ \rightarrow ll\nu \nu$).  The
four-lepton channel is more sensitive because of the very small
background levels, but the number of expected signal events is also
small.  The $ll\nu\nu$ final state, on the other hand, will have
larger numbers of expected signal events, but the larger backgrounds
make the channel less sensitive.

\begin{figure}
\includegraphics[width=0.48\textwidth]{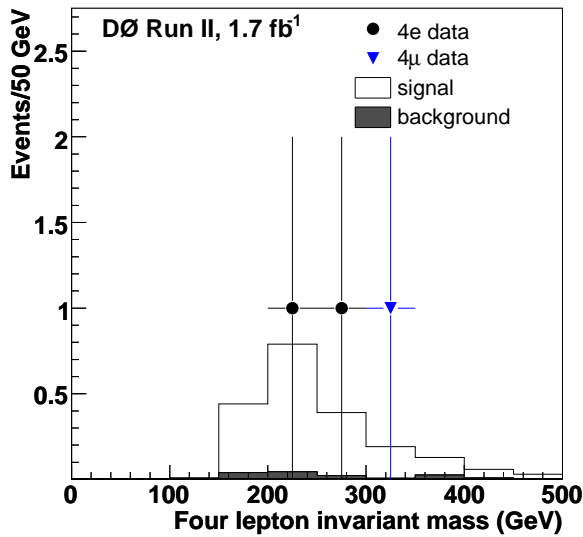}
\caption{\label{fig:ZZ_D0}Invariant mass of four-lepton events observed in $ZZ$ search.}
\end{figure}

The first observation of $ZZ$ production was reported by \DZero in 1.7
\fb of data~\cite{ZZ_D0}.  The search in the $ZZ \rightarrow llll$
channel was conducted in events with four electrons, four muons, or
two electrons and two muons.  Events with electrons were divided
further based on the number of electrons in the central region,
creating several categories with different expected
signal-to-background ratios.  The number of expected background events
was taken from Monte Carlo simulation.  Three four-lepton signal
events (two with four electrons and one with four muons) were
observed.  The expected background was $0.14^{+0.03}_{-0.02}$ events.
The invariant mass of the four leptons is shown in
Fig.~\ref{fig:ZZ_D0} superimposed on the expected shape of the $ZZ$
signal and the predicted background contribution.  This result was
combined with a less powerful search in $ZZ\rightarrow ll\nu\nu$,
resulting in an observation of the signal with a significance of
5.7$\sigma$.

CDF also presented strong evidence for $ZZ$ production in
1.9\fb~\cite{ZZ_CDF}.  In the $ZZ \rightarrow llll$ channel, three
events were observed with a predicted background of
$0.096^{+0.092}_{-0.063}$.  The search in the $ZZ \rightarrow
ll\nu\nu$ channel calculated event-by-event probability densities for
the $WW$ and $ZZ$ processes to discriminate between them.  Combination
of the $llll$ and $ll\nu \nu$ channels yielded a signal significance
of 4.4$\sigma$ and a cross section measurement of
$\sigma(p\bar{p}\rightarrow ZZ) = 1.4^{+0.7}_{-0.6}$ pb.

\subsection{Precise $WW$ Cross Section Measurement}

For ``high'' Higgs masses ($m_{H}>135$~GeV), the most sensitive
channel at the Tevatron is direct Higgs production with the Higgs
decaying to two $W$ bosons and the bosons subsequently decaying
leptoically ($H
\rightarrow WW \rightarrow l\nu l\nu$).  Measurement of the Standard Model
production of $WW \rightarrow l\nu l\nu$ events is a useful test of
our understanding of this final state.  It also provides a measurement
of the primary background to the Higgs search.

\begin{figure}
\includegraphics[width=0.48\textwidth]{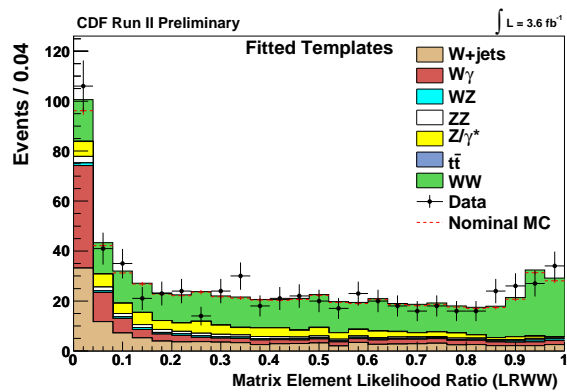}
\caption{\label{fig:WW_CDF}Fit to matrix element likelihood ratio used to extract the $WW$ cross section in the $l\nu l\nu$ final state.}
\end{figure}

Both experiments have recently published precise measurements of the
$WW$ cross section in the $lvlv$ mode.  In 3.6 \fb at CDF, events with
two opposite-sign leptons (electrons or muons) were selected.  The
primary backgrounds were $W$+jet, $W\gamma$, and Drell-Yan events; in
total roughly equal amounts of signal and background were expected.  A
matrix element technique was used, meaning the differential cross
sections of signal and several background processes were evaluated to
derive an event-by-event probability density.  A likelihood ratio
between probabilities was formed to discriminate between signal and
background.  A fit to this likelihood ratio is shown in
Figure~\ref{fig:WW_CDF}.  The extracted $WW$ cross section was
$\sigma(p\bar{p}) \rightarrow 12.1 \pm 0.9$ (stat)
$^{+1.6}_{-1.4}$(syst) pb, the most precise measurement of this
process at the Tevatron to date.

In 1.0 \fb at \DZero, the event selection was optimized to give the
lowest statistical and systematic uncertainties in each of three
lepton channels ($ee$, $e\mu$, and $\mu\mu$).  The predicted signal
and background event yields were then used to extract the cross
section.  The three channels were combined to give a cross section
measurement, $\sigma(p\bar{p} \rightarrow WW) = 11.5 \pm 2.1$
(stat+syst) $\pm 0.7$ (lumi) pb.

\section{Semi-leptonic Decay Channels}

For ``light'' Higgs masses ($m_{H}<135$~GeV, the most sensitive search
channels at the Tevatron are those where a Higgs is produced in
association with a $W$ or $Z$ boson with the $W$ or $Z$ decaying
leptonically and the Higgs decaying to $b\bar{b}$.  The $W$ or $Z$
leptonic decays with one or two identified leptons or large missing
transverse energy (from invisible decays or un-identified leptons) are
all used in the Higgs searches.  Studying the analogous final states
from semileptonic decays of $WW$, $WZ$, and $ZZ$ events can improve
our understanding of these channels.  The diboson results presented
here do not require $b$-tagging, which is an important difference with
respect to the low-mass Higgs searches.

Diboson events where one boson decays to two quarks ($WW/WZ
\rightarrow l\nu qq$, $ZW/ZZ \rightarrow \nu\nu qq$, and $ZW/ZZ
\rightarrow llqq$) suffer large
backgrounds from $W/Z+$jets events.  As a result measurements carried
out in these channels will be less precise than those from the fully
leptonic channels.  Recent Tevatron results have proven that it is
possible to observe these processes, both with multivariate techniques
similar to those used in Higgs searches and with simpler techniques
relying on the invariant mass of the two jets (dijet mass or
$M_{jj}$).

One measurement in the channel with large missing transvere energy and
two jets and three measurements with one identified lepton and two
jets are presented below.  A feature of all of them is that $W
\rightarrow qq\prime$ and $Z\rightarrow q\bar{q}$ are very challenging
to distinguish due to detector resolution effects, so the signals
measured are a sum of diboson production processes.

Table~\ref{tab:semi} summarizes the measurements performed in the
semi-leptonic decay modes.  They are described in more detail below.

\begin{table*}
\begin{center}
\caption{\label{tab:semi}Summary of measurements in semi-leptonic decay modes.}
\begin{tabular}{|c|c|c|c|c|}
\hline
\hline
 & & \multicolumn{3}{c|}{Cross section [pb]} \\
Process and channel & Analysis technique & CDF & \DZero & NLO prediction \\
\hline
$WW+WZ+ZZ \rightarrow \met jj$ & $M_{jj}$ fit & 18.0 $\pm$ 3.8 & & 16.8 $\pm$ 0.8 \\
\hline
$WW+WZ \rightarrow l \nu jj$ & $M_{jj}$ fit & 14.4 $\pm$ 3.8 & 18.5 $\pm$ 5.7 & 15.4 $\pm$ 0.8 \\
$WW+WZ \rightarrow l \nu jj$ & Random Forest classifier & & 20.2 $\pm$ 4.5 & 15.4 $\pm$ 0.8 \\
$WW+WZ \rightarrow l \nu jj$ & Matrix elements & 17.7 $\pm$ 3.9 & & 15.4 $\pm$ 0.8 \\
\hline
\hline
\end{tabular}
\end{center}
\end{table*}

\subsection{$WW+WZ+ZZ \rightarrow jj$ at CDF}

CDF reported the first observation of diboson production where one
boson decays to leptons and the other to hadrons~\cite{METjets_CDF} in
3.5 \fb.  Events with very large missing transverse energy ($\met>$60
GeV) and exactly two jets were used for this observation.  No veto on
events with a lepton in the final state was imposed.  The analysis was
therefore sensitive to a sum of $WW, WZ$, and $ZZ$ processes.

One challenge in this channel was understanding the QCD multi-jet
background (MJB).  An event with many jets can have large fake $\met$
because of mismeasurement of the jet energies, often stemming from
instrumental effects, making this background difficult to model. The
size of the MJB background was significantly reduced by imposing cuts
on the $\met$ significance and the angle between the $\met$ and the
jets.  The remaining MJB was modeled with a data sample enriched in
multi-jet events, selected by finding events with a large difference
in direction between their track-based missing transverse momentum and
their calorimeter-based missing transvere energy.

The second large background stemmed from electroweak processes
($W$+jets, $Z$+jets, and $t$-quark production).  These backgrounds
were modeled using Monte Carlo.  The uncertainty on the model was
evaluated with data, using $\gamma+$jet events.

\begin{figure}
\includegraphics[width=0.48\textwidth]{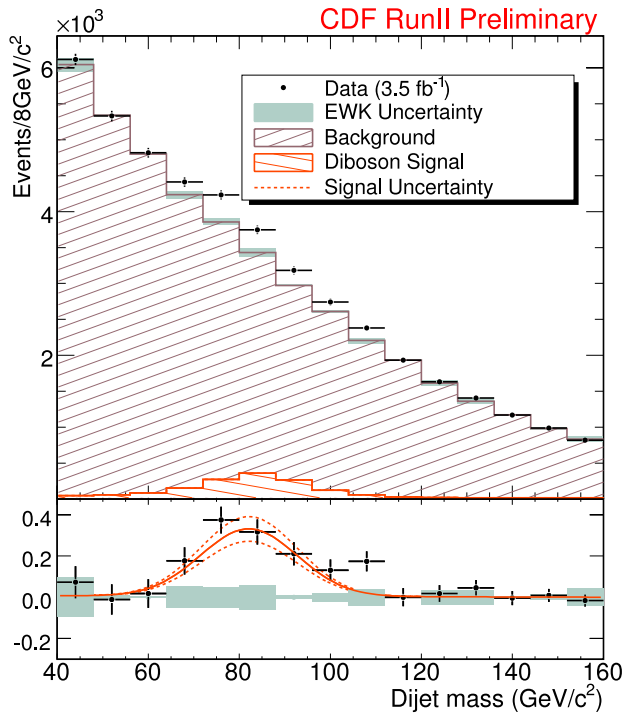}
\caption{\label{fig:METjets_CDF}Fit to mass dijet mass distribution to extract $WW+WZ+ZZ$ cross section in events with large missing transverse energy and two jets.}
\end{figure}

The signal cross section was extracted by a fit to the dijet mass
spectrum with signal and background templates.  The fit is shown in
Figure~\ref{fig:METjets_CDF}.  The fitted cross section was found to
be $\sigma(p\bar{p} \rightarrow WW+WZ+ZZ) = 18.0 \pm
2.8$(stat)$\pm2.4$(syst)$\pm1.1$(lumi), with the dominant systematic
uncertainty due to the jet energy scale.  The signal was observed with
a significance of 5.3$\sigma$.

\subsection{$WW+WZ \rightarrow l \nu j j$ at \DZero}

The first evidence of $WW+WZ \rightarrow l\nu jj$ was reported by the
\DZero collaboration~\cite{WWWZ_D0} in 1.07 \fb.  Events with a
well-identified electron or muon, at least two jets, and $\met>$20 GeV
were selected.  The background due to QCD multi-jet events was reduced
by requiring the transverse mass of the lepton-$\met$ system to be
larger than 35 GeV.  The remaining backgrounds were dominated by
$W$+jets production, with some QCD multi-jet, $Z$+jets, and top quark
production contributing as well.

The QCD multijet background was modeled using data with somewhat
loosened lepton requirements.  The $W$+jets background was modeled
with simulated events from Algpen interfaced with the Pythia parton
shower.  The modeling of the $W$+jets background was critical to the
analysis, so careful comparison between data and Monte Carlo was
carried out.  Discrepancies in the jet $\eta$ distributions and the
$\Delta R_{jj}$ distributions were observed; the models were
reweighted to agree with data.

Once confident in the modeling, a Random Forest Classifier (RF) was
used to discriminate between signal and background events.  Several
kinematic variables, such as the dijet mass, were used as inputs and
the RF was trained on part of the background to build a classification
for each event.

\begin{figure}
\includegraphics[width=0.48\textwidth]{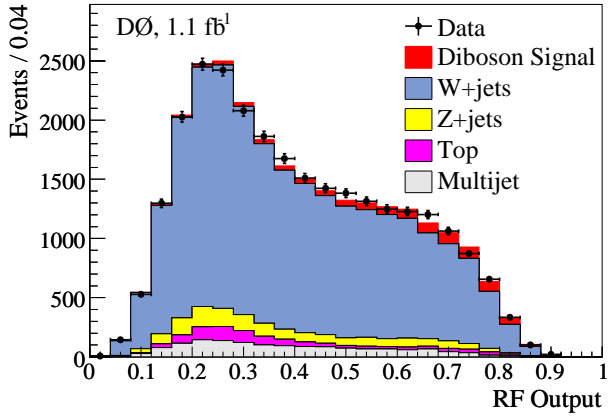}
\caption{\label{fig:WWWZ_D0}Fit to Random Forest Classifier output used to extract the $WW+WZ \rightarrow l\nu jj$ cross section.}
\end{figure}

The distribution of the RF output observed in data was fitted to a sum
of predicted RF templates to extract the signal significance and cross
section.  The fit is shown in Figure~\ref{fig:WWWZ_D0}.  The measured
cross section is $\sigma(p\bar{p} \rightarrow WW+WZ) = 20.2 \pm 4.5$
pb, with dominant systematic uncertainties from the modeling of the
$W$+jets background distribution and the jet energy scale.  The
significance of the observed signal is 4.4$\sigma$.

\begin{figure}
\includegraphics[width=0.48\textwidth]{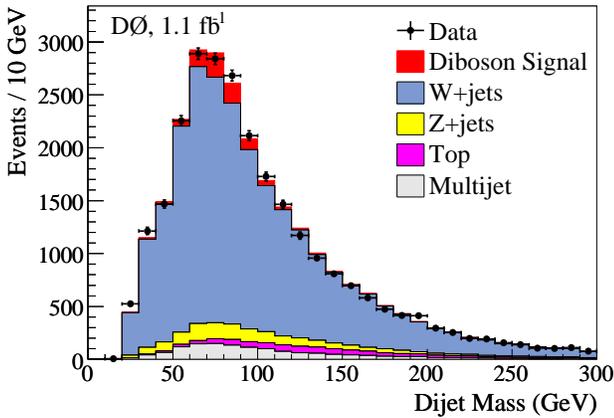}
\caption{\label{fig:WWWZ_D0_Mjj}Dijet mass distribution for data, backgrounds, and expected $WW+WZ$ signal in $l\nu jj$ final state.}
\end{figure}

The same analysis was carried out using only the dijet mass
distribution rather than the random forest classifier output.  Since
less information about the event is used, the measurement is expected
to be less precise.  The dijet mass is shown in
Figure~\ref{fig:WWWZ_D0_Mjj}.  The result of the fit was
$\sigma(p\bar{p} \rightarrow WW+WZ) = 18.5 \pm 5.7$, compatible with
the result from the RF classifier.

\subsection{$WW+WZ \rightarrow l\nu jj$ at CDF}

The first observation of $WW+WZ \rightarrow l\nu jj$ was presented by
CDF in 2.7 \fb.  Events with an isolated electron or muon, exactly two
jets, and $\met>$20 GeV were chosen.  Strong cuts were imposed in
events with an electron to reduce the QCD multi-jet background due to
jets faking electrons.  As a result the measurement was dominated by
events with muons.

Validation of the background modeling was also critical for this
analysis.  Three regions were chosen according to the dijet mass: the
signal-rich region with $55<M_{jj}<120$ GeV and two control regions
with $M_{jj}<55$ GeV and $M_{jj}>120$ GeV where very little signal was
expected.  Good modeling was observed in each region.  Some
mismodeling in the dijet mass was observed when the control regions
were combined, and a corresponding systematic uncertainty on the shape
of the $W$+jets background was applied.

\begin{figure}
\includegraphics[width=0.48\textwidth]{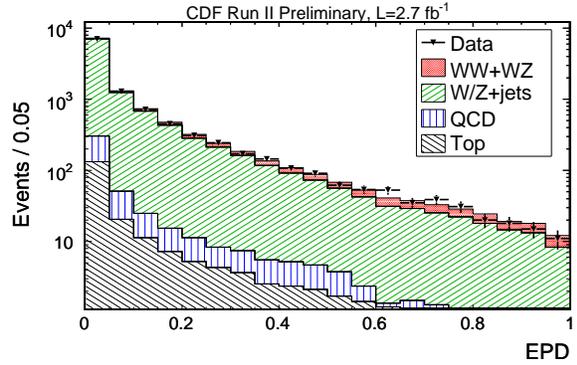}
\caption{\label{fig:WWWZ_CDF}Distribution of the discriminant derived from matrix elements used to extract the $WW+WZ$ cross section in the $l \nu jj$ final state}
\end{figure}

Matrix element calulations were used to discriminate between signal
and background.  The differential cross sections of signal and
background processes were evaluated for each event.  A discriminant
called the Event Probability Discrminant was formed to separate signal
from background.  The predicted shapes of signal and background
discriminants were fit to the data to extract the diboson cross
section.  The data superimposed on the background templates is shown
in Figure~\ref{fig:WWWZ_CDF}.  The measured cross section is $17.7 \pm
3.9$ pb where the dominant systematic uncertainty was the jet energy
scale.  The significance of the signal observation was 5.4$\sigma$.

\begin{figure}
\includegraphics[width=0.48\textwidth]{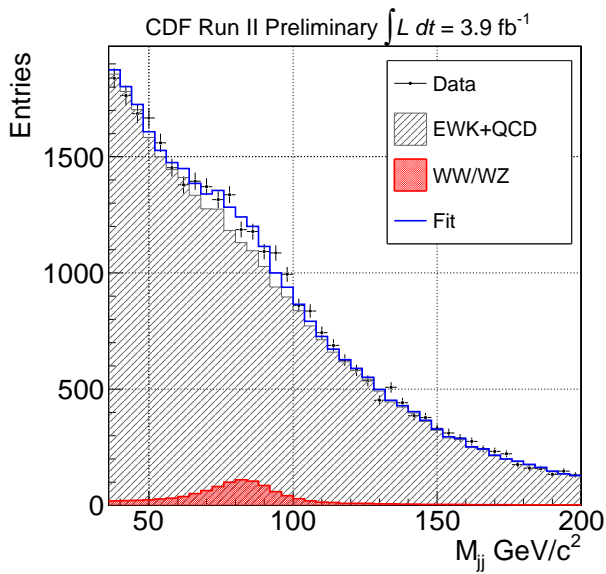}
\caption{\label{fig:WWWZ_Mjj}Distribution of the dijet mass for data and signal and background models for $WW+WZ \rightarrow l\nu jj$.}
\end{figure}

A second complimentary search was carried out at CDF using a larger
data sample of 3.9 \fb by fitting the $M_{jj}$ spectrum.  The event
selection criteria were adjusted to achieve a smoothly falling shape
in the $M_{jj}$ distribution of the backgrounds.  In particular, the
$p_{T}$ threshold on each individual jet was lowered, but the $p_{T}$
of the dijet system was required to be larger than 40 GeV.  The
diboson signal resulted in a bump on top of the background, as shown
in Figure~\ref{fig:WWWZ_Mjj}.  A fit with signal and background
templates was carried out, and the extracted cross section was $14.4
\pm 3.1$(stat) $\pm 2.2$(sys) pb, corresponding to an observed
significance of 4.6$\sigma$.

\section{Conclusions}

Several measurements of diboson production cross sections have been
carried out recently at the Tevatron.  These measurements can be a
testing ground for techniques used in searches for the Standard Model
Higgs boson.  Measurements are performed in many different final
states, ranging from those with several identified leptons to those
with no identified leptons and two jets.  Different analysis
techniques are also used, from counting signal events to performing
fits to kinematic quantities to techniques involving classifiers or
matrix element calculations.  There is good agreement between results
found at CDF and \DZero, as well as good agreement with NLO
predictions.

\bigskip
\begin{acknowledgements}
We thank the Fermilab staff and the technical staffs of the
participating institutions for their vital contributions. This work
was supported by the U.S. Department of Energy and National Science
Foundation; the Italian Istituto Nazionale di Fisica Nucleare; the
Ministry of Education, Culture, Sports, Science and Technology of
Japan; the Natural Sciences and Engineering Research Council of
Canada; the National Science Council of the Republic of China; the
Swiss National Science Foundation; the A.P. Sloan Foundation; the
Bundesministerium f\"ur Bildung und Forschung, Germany; the World
Class University Program, the National Research Foundation of Korea;
the Science and Technology Facilities Council and the Royal Society,
UK; the Institut National de Physique Nucleaire et Physique des
Particules/CNRS; the Russian Foundation for Basic Research; the
Ministerio de Ciencia e Innovaci\'{o}n, and Programa
Consolider-Ingenio 2010, Spain; the Slovak R\&D Agency; and the
Academy of Finland.
\end{acknowledgements}

\end{document}